\documentclass{article}
 \usepackage{amssymb}
\usepackage[dvips]{graphicx}

\newcommand{\bc}{\begin{center}}
\newcommand{\ec}{\end{center}}
\newcommand{\ba}{\begin{array}}
\newcommand{\ea}{\end{array}}
\newcommand{\beq}{\begin{equation}}
\newcommand{\eeq}{\end{equation}}
\newcommand{\beqa}{\begin{eqnarray}}
\newcommand{\eeqa}{\end{eqnarray}}
\newcommand{\no}{\nonumber}
\def\mato#1{\left(\ba{#1}}
\def\matf{\ea\right)}

\def\sect#1{\section{#1}\setcounter{equation}{0}}
\def\ssect#1{\subsection{#1}}

\def\eq#1{(\ref{#1})}
\def\lab#1{\label{#1}}

\def\Psic{\Psi_{pol}}
\def\bPsic{\b{\Psi}_{pol}}
\def\PsicR{(\Psi_{pol})_R}
\def\PsicL{(\Psi_{pol})_L}
\def\PsicdR{(\Psi_{pol})_R^\dag}
\def\PsicdL{(\Psi_{pol})_L^\dag}
\def\U{{\cal U}}

\def\End{{\rm End}}
\def\Or{{\cal O}}
\def\Hperp{{\cal H}}
\def\piperp{\pi_{\cal H}}

\def\iK{{\bf K}}
\def\genK{K}

\def\d{\partial}
\def\frc#1#2{\frac{#1}{#2}}
\def\Tr{\mbox{Tr}}
\def\ep{\epsilon}
\def\b#1{\bar{#1}}
\def\om{\omega}
\def\ga{\gamma}
\def\lt#1{\left#1}
\def\rt#1{\right#1}
\def\t#1{\tilde{#1}}
\def\rx{{\rm x}}
\def\ry{{\rm y}}
\def\T{{\cal T}}
\def\h#1{\hat{#1}}
\def\G{\Gamma}
\def\<{\langle}
\def\>{\rangle}
\def\arctanh{{\rm arctanh}}

\def\a{\alpha}
\def\l{\lambda}

\def\dist{{\rm d}}

\def\N{{\mathbb N}}
\def\R{{\mathbb R}}

\begin{document}

\pagestyle{empty}
\pagenumbering{arabic}

\hfill RUNHETC-2003-08
\vspace{3cm}
\bc
\Large\bf
Two-point correlation functions of scaling fields\\
in the Dirac theory on the Poincar\'e disk
\ec
\vspace{2cm}
\bc
Benjamin Doyon

NHETC, Department of Physics and Astronomy\\
Rutgers University\\
Piscataway, NJ 08854, USA
\ec
\vspace{2cm}
\bc
\bf Abstract
\ec

A result from Palmer, Beatty and Tracy suggests that the two-point
function of certain spinless scaling fields in a free Dirac theory
on the Poincar\'e disk can be described in terms of Painlev\'e VI
transcendents. We complete and verify this description by fixing
the integration constants in the Painlev\'e VI transcendent
describing the two-point function, and by calculating directly in
a Dirac theory on the Poincar\'e disk the long distance expansion
of this two-point function and the relative normalization of its
long and short distance asymptotics. The long distance expansion
is obtained by developing the curved-space analogue of a form
factor expansion, and the relative normalization is obtained by
calculating the one-point function of the scaling fields in
question. The long distance expansion in fact provides part of the
solution to the connection problem associated with the Painlev\'e
VI equation involved. Calculations are done using the formalism of
angular quantization. \vfill\noindent April 2003

\newpage
\pagestyle{plain}
\setcounter{page}{1}

\sect{Introduction}

Quantum Field Theory (QFT) in curved space-time is a subject of
great interest which has been studied from many viewpoints (see
for instance \cite{BirrelDavies}). The main goal of QFT is the
reconstruction of a set of correlation functions in which is
embedded the physical information provided by the theory. Such a
reconstruction is an important problem for QFT on curved
space-time, where the structure of the theory is more subtle. A
simple but non-trivial curved space-time with Euclidean signature
is the Poincar\'e disk. It is maximally symmetric, which allows
some techniques on two-dimensional flat space to be generalized to
this space, and has a negative Gaussian curvature. The effect of a
negative curvature is interesting to study: as was argued in
\cite{CW}, a negative curvature can be used as an infrared
regulator for Euclidean QFT.

In two-dimensional flat space, a class of non-trivial correlation
functions can be studied by their relation to the problem of
isomonodromic deformations \cite{SMJ}. Correlators of certain
scaling fields in a free massive Dirac theory are tau functions
\cite{SMJ, JMU, JM, P} for the isomonodromic deformation problem
associated to some Painlev\'e equations: they can be expressed in
terms of Painlev\'e transcendents. The scaling fields in question,
$\Or_\alpha=\Or_{-\a}^\dag ,\,-1<\alpha<1$, are spinless,
$U(1)$-neutral and have scaling dimension $\alpha^2$. They are not
mutually local with respect to the Dirac field. Their mutual
locality index with the Dirac field is $\alpha$, that is, the
Dirac field $\Psi$ takes a factor, $\Psi \to e^{2\pi i \a} \Psi$,
when continued counterclockwise around the field $\Or_\a$. Their
physical interest stems in particular from the fact that
correlators of some of these scaling fields are simply related to
the scaling limit of correlators of local variables in the Ising
model \cite{ZI,ST,WMTB}.

In \cite{PBT}, the authors generalized the method of isomonodromic
deformation to the study of determinants of the Dirac operator on
the Poincar\'e disk. Their results suggest that the two-point
function of scaling fields $\Or_\a$ in a Dirac theory on the
Poincar\'e disk can be expressed in terms of Painlev\'e VI
transcendents. The general form of the Painlev\'e VI differential
equation is: \beqa &&  w'' - \frc12\lt( \frc1w + \frc1{w-1} +
\frc1{w-s} \rt) (w')^2 + \lt(\frc1s + \frc1{s-1} +
\frc1{w-s}\rt)w' \no\\&& = \frc{w(w-1)(w-s)}{s^2(1-s)^2} \lt(
\frc{(1-4\mu^2)s(s-1)}{2(w-s)^2} - \frc{(\t\lambda-1)^2s}{2w^2}
        + \frc{\gamma(s-1)}{(w-1)^2} + \frc{\lambda^2}2 \rt)
\lab{painleveintro}\eeqa where $w=w(s)$ and the primes mean
derivatives with respect to $s$. There are four parameters:
$\mu,\,\gamma,\,\lambda,\,\t\lambda$. For the description of the
two-point function $\<\Or_\a(x) \Or_{\a'}(y)\>$ in the free Dirac
theory on the Poincar\'e disk, with fermion mass $m$ and Gaussian
curvature $-\frc1{R^2}$, the parameters are fixed to $\mu=mR$,
$\lambda=\a-\a'$, $\t\lambda=\a+\a'$ and $\ga=0$. The two-point
function is then identified with the associated tau function
$\tau(s)$, with $s$ simply related to the geodesic distance
$\dist(x,y)$ between the points $x$ and $y$:
$$
    \<\Or_\a(x) \Or_{\a'}(y)\>=\tau(s),\ \ \  s =\tanh^2\lt( \frc{\dist(x,y)}{2R} \rt).
$$
Up to normalization, the tau function is given by \cite{PBT}:
\beqa\lab{tau}
    \frc{d}{ds} \ln\tau(s) &=& \frc{s(1-s)}{4w(1-w)(w-s)} \lt( w'-\frc{1-w}{1-s} \rt)^2 - \frc{\mu^2}{w-s} \no\\
    &&  + \frc{{\t\lambda}^2}{4(1-s)w} + \frc{\lambda^2 w}{4s(1-s)} - \frc{\lambda^2}{4s}
        + \frc{4\mu^2-{\t\lambda}^2 - \lambda^2}{4(1-s)} .
\eeqa

This description for the correlator needs to be completed and to
be compared with direct calculation in the Dirac theory on the
Poincar\'e disk. In order to complete it, one must supply
appropriate integration constants specifying the Painlev\'e
transcendent that describes the correlator. First, the exponent
$2\a\a'$ in the short distance power law of the two-point
function, \beq\lab{norm}
    \<\Or_\a(x) \Or_{\a'}(y)\> \sim \<\Or_{\a+\a'}\> \dist(x,y)^{2\a\a'}
    \ \mbox{ as } x\to y \;,
\eeq
specifies the exponent in the asymptotic behavior of $w(s)$ near the critical point $s=0$:
\beq\lab{ws0}
    w \sim Bs^{\a+\a'} \ \mbox{ as } s\to0,
\eeq
where $B$ is some constant and where we must have $0<\a+\a'<1$.
Second, the cluster property of the two-point function,
\beq\lab{leadinglongdist}
    \<\Or_\a(x) \Or_{\a'}(y)\> = \<\Or_\a\>\<\Or_{\a'}\> F\lt(\frc{\dist(x,y)}{2R}\rt) ,
    \ \ \
    \lim_{t \to \infty} F\lt(t\rt) = 1,
\eeq
specifies, for $\mu>\frc12$, the exponent in the asymptotic behavior of $w(s)$ near the critical point $s=1$:
\beq\lab{ws1}
    1-w \sim A (1-s)^{1+2\mu} \ \mbox{ as } s\to 1,
\eeq where again, $A$ is some constant.

One can expect that the exponents in \eq{ws0} and \eq{ws1} form a
set of integration constants fixing the Painlev\'e transcendent.
However, one doesn't know {\em a priori} that there exists a
solution to the Painlev\'e equation with both behaviors \eq{ws0}
and \eq{ws1}. In addition, even if such a solution exists, this
set is not the most convenient. One cannot use it for instance to
provide initial conditions for numerically solving the
differential equation \eq{painleveintro}. It is more appropriate
to fix the full expansion of the Painlev\'e transcendent near the
singular point $s=1$, that is, to specify the constant $A$ in
\eq{ws1}. Fixing this constant is part of the solution to the
connection problem for the particular Painlev\'e VI equation that
we are considering, that is, the problem of relating the behaviors
of the Painlev\'e transcendent near its various critical points.
The expansion of the Painlev\'e transcendent near $s=1$ is
directly related to the long distance expansion of the two-point
function: \beq\lab{Flarged}
    F\lt(t\rt) = 1 -
    A \frc{(\mu+\a)(\mu+\a')}{(2\mu+1)^2} 4^{2\mu+1} e^{-(4\mu+2)t} +
    O\lt( e^{-(4\mu+4)t}\rt).
\eeq To our knowledge, the theory of Painlev\'e VI equations
\eq{painleveintro} in the case $\ga=0$ provides no expression for
$A$ in terms of the exponents in \eq{ws0} and \eq{ws1}. We
calculated $A$ by evaluating directly in a free massive Dirac
theory on the Poincar\'e disk the long distance expansion of the
two-point function: \beq\lab{A}
    A = \frc{\sin(\pi\a) \sin(\pi\a') \G(\mu+\a) \G(1+\mu-\a)
    \G(\mu+\a') \G(1+\mu-\a') }{
    \pi^2 \G(1+2\mu)^2}
\eeq The asymptotics \eq{ws0} can also serve as initial condition
once $B$ is known, although it is numerically not as efficient.
From results by Jimbo concerning Painlev\'e VI equations \cite{J},
one can calculate the constant $B$ (see section 7) -- this is
another part of the solution to the connection problem.

Another important quantity concerning the two-point function that
is not provided by the theory of Painlev\'e VI equations is an
expression relating the normalization of the leading long distance
asymptotics of the two-point function to that of its leading short
distance asymptotics. Comparing \eq{norm} and
\eq{leadinglongdist}, one sees that the relation between
normalizations can be obtained by calculating the one-point
function $\<\Or_\a\>$. Notice that the condition \eq{norm} fixes
the normalization of the scaling fields $\Or_\a$. This
normalization being fixed, the one-point function $\<\Or_\a\>$ is
unambiguous. By a simple generalization of calculations done on
flat space in \cite{AlZ, LZ}, we found
$$
    \<\Or_\a\> = (2R)^{-\a^2} \prod_{n=1}^{\infty} \lt( \frc{1-\frc{\a^2}{(\mu+n)^2}}{1-\frc{\a^2}{n^2}}\rt)^n.
$$

The long distance asymptotics of the two-point function is
obtained by a ``form factor'' expansion, that is, an expansion
obtained by inserting between the two operators in an appropriate
vacuum expectation value a resolution of the identity on an
appropriate Hilbert space. Actual calculations of form factors on
this Hilbert space are done in the formalism of angular
quantization. More precisely, matrix elements of operators
corresponding to the scaling fields $\Or_\a$ between the vacuum
and an excited state are evaluated through appropriate traces on
the Hilbert space of angular quantization, following ideas from
\cite{L1, BL, KLP}. The one-point function $\<\Or_\a\>$ is
similarly evaluated through appropriate traces on the Hilbert
space of angular quantization.

The plan of the paper is as follows. In section 2 we briefly
recall standard results concerning the free Dirac fermion on the
Poincar\'e disk. In section 3 we express the long distance
expansion of the two-point function as a form factor expansion
using canonical quantization. In section 4 we evaluate, up to
normalization, the form factors of the scaling fields $\Or_\a$ by
calculating appropriate traces on the Hilbert space of angular
quantization, and in section 5 we compute the one-point function
$\<\Or_\a\>$, which specifies the normalization of the form
factors. In section 6 we discuss the long distance expansion of
the two-point function, and finally in section 7 we briefly
elaborate on the description of \cite{PBT} for the two-point
function in terms of a Painlev\'e VI transcendent and clarify our
results in this context.

\sect{Free Dirac fermion on the Poincar\'e disk}

With appropriate complex coordinates $z={\rm x} + i{\rm y},\,\b{z}={\rm x} - i {\rm y}$, the
Poincar\'e disk can be brought to the region $|z|<1$ in the complex $z$--plane. For a Gaussian curvature
$-\frc1{R^2}$ the metric is then specified by
$$
    ds^2 = \frc{dzd\b{z} (2R)^2}{(1-|z|^2)^2} .
$$
The metric is $SU(1,1)$ invariant. The transformation of coordinates corresponding to the group element
$g = \mato{cc} a & b \\ \b{b} & \b{a} \matf ,\; |a|^2-|b|^2 = 1$ is
$$
    z \mapsto \frc{az + \b{b}}{bz + \b{a}} \;,\; \b{z} \mapsto \frc{\b{a}\b{z} + b}{\b{b}\b{z} + a} .
$$
The geodesic distance $\dist(x_1,x_2)$ between points $x_1=(z_1,\b{z}_1)$ and $x_2=(z_2,\b{z}_2)$ on
the Poincar\'e disk is given by
\beq\lab{geod}
    \dist(x_1,x_2) = 2R\,{\arctanh} \lt( \frc{|z_1-z_2|}{|1-z_1\b{z}_2|} \rt).
\eeq

The free massive Dirac action with fermion mass $m$ in this system of coordinates is
\beq\lab{theory}
    {\cal A} = \int d{\rm x} d{\rm y} \, \b\Psi \lt(\ga^\rx\d_\rx + \ga^\ry\d_\ry + \frc{2\mu}{1-|z|^2}\rt) \Psi,
\eeq
where as in the introduction $\mu=mR$, and $\Psi = \mato{c} \Psi_R \\ \Psi_L \matf$,
$\b\Psi = \Psi^\dag \ga^{\rm y}$. We choose the Dirac matrices as
$$
    \ga^{\rm x} = \mato{cc} 0 & i \\ -i & 0 \matf \;,\; \ga^{\rm y} = \mato{cc} 0 & 1 \\ 1 & 0 \matf.
$$
The two-point function that we are interested in can be represented by the appropriately regularized
Euclidean functional integral
\beq\lab{pint}
    \<\Or_\a(x) \Or_{\a'}(y)\> = \int_{{\cal F}_{\a,\a'}} [{\cal D}\Psi {\cal D}\b\Psi] e^{-{\cal A}}.
\eeq
The integration is over the space ${\cal F}_{\a,\a'}$ of field configurations vanishing on the boundary of the disk
and such that $\Psi,\,\b\Psi$ acquire phases when continued counterclockwise around the points $x$ and $y$:
\beq{\cal F}_{\a,\a'}: \lt\{\ba{l}
    \mbox{around }x: \  \Psi \to e^{2\pi i \a}\Psi,\ \b\Psi \to e^{-2\pi i \a} \b\Psi \\
    \mbox{around }y: \ \Psi \to e^{2\pi i \a'}\Psi,\ \b\Psi \to e^{-2\pi i \a'} \b\Psi
\ea\rt. .\lab{qp}\eeq

\sect{Hilbert space}

An expansion of the functional integral \eq{pint} for large
geodesic distance between the points $x$ and $y$ is most
conveniently obtained by using the operator formalism. We will
construct a Hilbert space $\Hperp$ on which this functional
integral is represented by a vacuum expectation value
$\<vac|\T[\Or_\a(x)\Or_{\a'}(y)]|vac\>$, where $|vac\>$ is a
$SU(1,1)$ invariant vacuum implementing vanishing boundary
conditions and where the fields $\Or_\a(x)$ and $\Or_{\a'}(y)$ act
on $\Hperp$ as operators implementing the quasi-periodicity
conditions \eq{qp}. The symbol $\T$ here denotes an appropriate
``time''-ordering, described below. The large distance expansion
will be obtained by inserting between the fields $\Or_\a(x)$ and
$\Or_{\a'}(y)$ a resolution of the identity in terms of a basis of
states that diagonalize a non-compact subgroup of the $SU(1,1)$
isometry group of the Poincar\'e disk.

\begin{figure}
\centering
\includegraphics{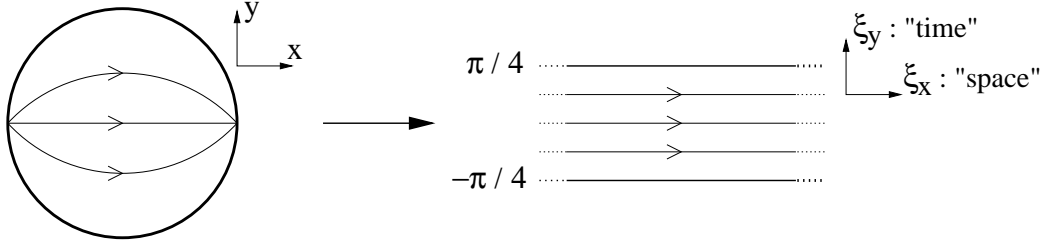}
\caption{mapping from the Poincar\'e disk to the strip. Lines with
an arrow represent orbits of $\iK$.} \label{fig1}
\end{figure}

We define the Hilbert space $\Hperp$ by quantizing the theory on
curves that are orbits of the non-compact subgroup $\iK$:
\beq\lab{isoK}
    \iK\,:\ g_q = \mato{cc} \cosh(q) & \sinh(q) \\ \sinh(q) & \cosh(q) \matf ,\; q\in \R.
\eeq Translations in the ``time'' direction, perpendicular to
these curves, are not isometries, so that the Hamiltonian is not
stationary. Translations in the ``space'' direction, along these
curves, are isometries, and a basis for $\Hperp$ will be obtained
by diagonalizing the generator of such translations. In order to
construct $\Hperp$, we first map the Poincar\'e disk onto the
strip (see Figure \ref{fig1}):
$$
    z=\tanh(\xi),\; \b{z} = \tanh(\b{\xi})
$$
with
$$
    \xi = \xi_\rx + i \xi_\ry ,\; \xi_\rx\in \R,\;
    -\pi/4<\xi_\ry<\pi/4.
$$
\noindent The action \eq{theory} for the fermion fields $\Psi_s =
\mato{c} (\Psi_s)_R \\ (\Psi_s)_L \matf$ on the strip is
$$
    {\cal A} = \int d\xi_{\rm x} d\xi_{\rm y} \, \b\Psi_s \lt(\ga^\rx\frc{\d}{\d\xi_\rx} + \ga^\ry\frc{\d}{\d\xi_\ry} + \frc{2\mu}{\cos(2\xi_\ry)}\rt) \Psi_s,
$$
where $\b\Psi_s = \Psi_s^\dag \ga^\ry$. Elements $g_q$ of the
subgroup $\iK$ are translations parallel to the strip: $\xi_\rx
\to \xi_\rx + q$, so that the quantization space is specified by
the curves $\xi_\ry=const.$ The coordinate $\xi_\ry$ is the
Euclidean ``time''. Correlation functions of local operators are
expressed as vacuum expectation values of ``time''-ordered
products, for instance
$$
    \<\Or_{\a_1}({\xi_\rx}_1,{\xi_\ry}_1)\Or_{\a_2}({\xi_\rx}_2,{\xi_\ry}_2)\cdots\> =
     \<vac|\T[\Or_{\a_1}({\xi_\rx}_1,{\xi_\ry}_1)\Or_{\a_2}({\xi_\rx}_2,{\xi_\ry}_2)\cdots]|vac\>,
$$
where the ``time''-ordering $\T$ means that operators have to be
ordered form left to right in decreasing values of their variable
$\xi_\ry$. The Hilbert space $\Hperp$ is a module for the
canonical equal-``time'' anti-commutation relations
$$
    \{\Psi_s(\xi_\rx,\xi_\ry),\Psi_s^\dag(\xi_\rx',\xi_\ry)\} = {\bf 1} \delta(\xi_\rx-\xi_\rx'),
$$
where now $\Psi_s(\xi_\rx,\xi_\ry)$,
$\Psi_s^\dag(\xi_\rx,\xi_\ry)$ are fermion operators acting on
$\Hperp$, related on different ``time'' slices $\xi_\ry=const$ by
the equations of motion. The vacuum state $|vac\>\in\Hperp$ is
$SU(1,1)$ invariant, and satisfies the conditions that the fermion
operators vanish on it at early ``times'' $\xi_\ry\to-\pi/4$ and
on its dual $\<vac|$ at late ``times'' $\xi_\ry\to\pi/4$:
\beq\lab{bcop} \ba{c} \displaystyle
    \lim_{\xi_\ry \to -\pi/4} \Psi_s(\xi_\rx,\xi_\ry)|vac\> =
    \lim_{\xi_\ry \to -\pi/4} \Psi_s^\dag(\xi_\rx,\xi_\ry)|vac\> = 0 \\
\displaystyle
    \lim_{\xi_\ry \to \pi/4} \<vac|\Psi_s(\xi_\rx,\xi_\ry) =
    \lim_{\xi_\ry \to \pi/4} \<vac|\Psi_s^\dag(\xi_\rx,\xi_\ry) = 0.
\ea\eeq

A basis for the Hilbert space $\Hperp$ diagonalizing the subgroup
$\iK$ is obtained by considering a module for appropriate modes
$A_\ep(\om)$, $A_\ep^\dag(\om)$ of the Fermion operators
$\Psi_s(\xi_\rx,\xi_\ry)$, $\Psi_s^\dag(\xi_\rx,\xi_\ry)$, where
$\ep=\pm$ represents the $U(1)$ charge. These modes appear in an
expansion in terms of partial waves satisfying the equations of
motion and diagonalizing the spin-$\frc12$ action of the subgroup
$\iK$: \beqa\lab{pwave}
    \Psi_s(\xi_\rx,\xi_\ry) &=& \int d\om\, \rho(\om) \, \lt(
        A_-^\dag(\om) e^{i\om \xi_\rx} \b{P}_-(\om,\xi_\ry) + A_+(\om) e^{-i\om \xi_\rx} P_+(\om,\xi_\ry) \rt)\\
    \Psi_s^\dag(\xi_\rx,\xi_\ry) &=& \int d\om\, \rho(\om) \, \lt(
        A_+^\dag(\om) e^{i\om \xi_\rx} \b{P}_+(\om,\xi_\ry) + A_-(\om) e^{-i\om\xi_\rx} P_-(\om,\xi_\ry) \rt) \no ,
\eeqa where the integrations are from $-\infty$ to $\infty$. We
choose the measure \beq\lab{meas}
    \rho(\om) = \frc{\G\lt(\frc12 + \mu + i\frc{\om}2\rt) \G\lt(\frc12 + \mu - i\frc{\om}2\rt)}{
            2\pi \G\lt(\frc12+\mu\rt)^2}
\eeq in order for the partial waves to be entire functions of the
spectral parameter $\om$; this analytical property will be used
later. The solutions for the partial waves can be written:
\beq\lab{pwaveexpr}
    P_+(\om,\xi_\ry) = \mato{c}
    2^{-2\mu-\frc12} e^{-i\frc{\pi}2\lt( \mu+\frc12-i\frc{\om}2\rt)}
        (1+e^{4i\xi_\ry})^\mu e^{\om \xi_\ry}
        F\lt(\mu,\frc12+\mu-i\frc{\om}2;1+2\mu;1+e^{4i\xi_\ry}\rt) \\
    2^{-2\mu-\frc12} e^{-i\frc{\pi}2\lt( \mu+\frc12+i\frc{\om}2\rt)}
        (1+e^{4i\xi_\ry})^\mu e^{-\om \xi_\ry}
        F\lt(\mu,\frc12+\mu+i\frc{\om}2;1+2\mu;1+e^{4i\xi_\ry}\rt)
    \matf
\eeq and \beq\lab{p2}
    \b{P}_\ep(\om,\xi_\ry) = (P_\ep(\om,-\xi_\ry))^\dag ,\;
    (P_-(\om,\xi_\ry))^t = \mato{cc} 1 & 0 \\ 0 & -1 \matf P_+(\om,\xi_\ry) ,
\eeq where ${\ }^t$ means transpose, and where $F(a,b;c;z)$ is
Gauss's hypergeometric function on a Riemann sheet delimited by
the branch cut $(-\infty,1]$, with $\lim_{z\to0,\;\Im m(z)>0}
F(a,b;c;z) = 1$. With this choice of solutions, the partial waves
$P_\pm(\om,\xi_\ry)$ vanish on the upper boundary $\xi_\ry =
\pi/4$ of the strip, and $\b{P}_\pm(\om,\xi_\ry)$ vanish on the
lower boundary $\xi_\ry = -\pi/4$ of the strip.

The canonical anti-commutation relations for the fermion operators
imply the following anti-commutation relations for the modes:
\beq\lab{Aac}
    \{A_{\ep}(\om), A^\dag_{\ep'}(\om')\} = \frc1{\rho(\om)} \delta(\om-\om') \delta_{\ep,\ep'}\,,\;
    \{A_{\ep}(\om), A_{\ep'}(\om')\} = \{A_{\ep}^\dag(\om), A_{\ep'}^\dag(\om')\} = 0 .
\eeq The Hilbert space is the Fock space over this algebra, and
the asymptotic conditions \eq{bcop} specify the vacuum $|vac\>$:
$$
    A_\pm(\om) |vac\> = 0
$$
which we normalize to $\<vac|vac\>=1$. A complete basis is given
by \beq\lab{basis}
    |\om_1,\ldots,\om_n\>_{\ep_1,\ldots,\ep_n} =
    A^\dag_{\ep_1}(\om_1) \cdots A^\dag_{\ep_n}(\om_n) |vac\>
\eeq for a given ordering of the $\om_j$'s, for instance the
``in-ordering'' $\om_1<\cdots<\om_n$.

The states constructed diagonalize the subgroup $\iK$:
$$
    \h{g}_q |\om_1,\ldots,\om_n\>_{\ep_1,\ldots,\ep_n} = e^{-iq(\om_1+\cdots+\om_n)}
    |\om_1,\ldots,\om_n\>_{\ep_1,\ldots,\ep_n}
$$
as well as the $U(1)$ charge, with eigenvalue
$\ep_1+\cdots+\ep_n$. Here and below we use the notation $\h{g}$
for representing the action of the group element $g$ on the space
$\Hperp$. These states can be used to obtain a resolution of the
identity on $\Hperp$:
$$
    {\bf 1}_{\Hperp} = \sum_{n=0}^{\infty} \frc1{n!} \sum_{\ep_1,\ldots,\ep_n} \int \lt(
    \prod_{j=1}^n d\om_j\,\rho(\om_j)\rt) \, |\om_1,\ldots,\om_n\>_{\ep_1,\ldots,\ep_n}
    {\ }_{\ep_n,\ldots,\ep_1}\<\om_n,\ldots,\om_1|,
$$
where states with different orderings of $\om_j$'s than the
in-ordering $\om_1<\cdots<\om_n$ are given by the same expression
\eq{basis} in terms of modes. They differ from states with
in-ordering by a sign through the anti-commutation relations
\eq{Aac}.

On the Hilbert space $\Hperp$, the scaling fields $\Or_\a(x)$ act
as appropriately regularized exponentials of line integrals of the
$U(1)$ current: \beq\lab{scalingfieldPsi}
    \Or_\a(x) \mapsto \piperp\lt(\exp\lt[\,
        2 \pi i \a \int_{{\cal C}_x} dx^\mu\ep_{\mu,\nu} \b\Psi  \ga^\nu \Psi
    \,\rt]\rt),
\eeq where ${\cal C}_x$ is a path from the position $x$ to the
boundary of the disk and $\piperp$ is the representation map on
the space $\Hperp$ (for a precise definition of the action of
similar scaling fields on the standard Hilbert space of a Dirac
theory on flat background, see for instance \cite{ST}). The
resolution of the identity on $\Hperp$ then gives the long
geodesic distance expansion of the two-point function $\<\Or_\a(x)
\Or_{\a'}(y)\> = \<vac|\T[\Or_\a(x)\Or_{\a'}(y)]|vac\>$. Using the
fact that the one-point function $\<\Or_\a\> = \<vac|\Or_\a|vac\>$
is non-zero and defining the function \beq\lab{Fff}
    F_\a(\om_1,\ldots,\om_n)_{\ep_1,\ldots,\ep_n} \equiv
    \frc{\<vac|\Or_\a(0)|\om_1,\ldots,\om_n\>_{\ep_1,\ldots,\ep_n}}{\<vac|\Or_\a|vac\>},
\eeq we have \beqa\lab{genexp}
    \<\Or_\a(x)\Or_{\a'}(y)\> &=& \<\Or_\a\> \<\Or_{\a'}\>
    \sum_{n=0}^{\infty} \frc1{n!} \sum_{\ep_1,\ldots,\ep_n} \int
    \lt(\prod_{j=1}^n d\omega_j \rho(\om_j)\rt) \;\times \\
&&\times\;  F_\a(\om_1,\ldots,\om_n)_{\ep_1,\ldots,\ep_n}
        (F_{-\a'}(\om_1,\ldots,\om_n)_{\ep_1,\ldots,\ep_n})^*
    e^{-i(\om_1+\cdots+\om_n) \frc{\dist(x,y)}{2R}} \no.
\eeqa In order to obtain this formula, one first brings $y$ to the
origin and $x$ to the real axis inside the Poincar\'e disk; this
can always be done by $SU(1,1)$ invariance of the correlator.
Since the subgroup $\iK$ generates geodesic translations along the
real axis inside the Poincar\'e disk, matrix elements of the
operator $\Or_\a(x)$ are related to those of $\Or_\a(0)$ by an
exponential factor involving the geodesic distance $\dist(x,0)$.
Using $SU(1,1)$ invariance again, one can replace this by
$\dist(x,y)$ for the correlator of fields at arbitrary points $x$
and $y$. In the next section we calculate the matrix elements
$\<vac|\Or_\a(0)|\om_1,\ldots,\om_n\>_{\ep_1,\ldots,\ep_n}$, which
we call ``form factors'' of the scaling fields $\Or_\a(0)$.

\sect{Angular quantization and form factors}

In order to construct form factors of local fields in quantum
integrable models, in \cite{L1,BL} the authors used the idea of
embedding the Hilbert space $\cal H$ of a quantum field theory in
two-dimensional flat space-time into a tensor product
$$
    {\cal H}_A \otimes {\cal H}_A^*
$$
of the Hilbert space of angular quantization ${\cal H}_A$ and its
dual ${\cal H}_A^*$. In angular quantization, the Hamiltonian is
taken as the generator of rotations around a given point. Form
factors of local fields can then be constructed as traces on the
angular Hilbert space ${\cal H}_A$. We will make a similar
construction in order to obtain the matrix elements
$\<vac|\Or_\a(0)|\om_1,\ldots,\om_n\>_{\ep_1,\ldots,\ep_n}$. The
angular Hamiltonian will be taken as the generator of rotations
around the center of the Poincar\'e disk; it is the generator for
the compact subgroup of the $SU(1,1)$ isometry group of the
Poincar\'e disk.

We first briefly develop the formalism of angular quantization
\cite{LZ,BL,KLP} for our theory. Angular quantization is done in
conformal polar coordinates $(\eta,\theta)$, where $\theta$ is the
Euclidean ``time'' and $-\infty < \eta <0$: \beq\lab{cylcoord}
    z=e^{\eta+i\theta},\; \b{z} = e^{\eta-i\theta}.
\eeq The fermion fields $\Psic = \mato{c} \PsicR \\ \PsicL \matf$
in these coordinates enter the Dirac action as \beq\lab{cylact}
    {\cal A} = \int_0^{2\pi} d\theta \int_{-\infty}^0 d\eta \,\bPsic \lt( \ga^\eta \d_\eta + \ga^\theta \d_\theta
    - \frc{\mu}{\sinh(\eta)} \rt) \Psic,
\eeq where $\bPsic = \Psic^\dag\ga^\theta$ and
$\ga^\eta,\,\ga^\theta$ are, respectively, the same matrices as
$\ga^\rx,\,\ga^\ry$. The angular Hamiltonian derived from this
action is \beq\lab{hamilt}
    H_A = \int_{-\infty}^0 d\eta :\Psic^\dag \ga^\theta \lt(\ga^\eta\d_\eta - \frc{\mu}{\sinh(\eta)}\rt) \Psic: ,
\eeq where $\Psic(\eta)$ and $\Psic^\dag(\eta)$ are now operators
on the angular Hilbert space ${\cal H}_A$ satisfying the canonical
anti-commutation relation
\[
    \{\Psic(\eta),\Psic^\dag(\eta')\} = {\bf 1} \delta(\eta-\eta') .
\]

The $\eta$-dependent mass term in the Hamiltonian \eq{hamilt}
produces a ``mass barrier'' effect somewhat similar to the effect
of the mass term in the theory on flat space \cite{LZ}; it
prevents the fermions from approaching too much the boundary of
the disk. More precisely, it imposes vanishing asymptotic
conditions for the fermion fields when $\mu>\frc12$ (such
asymptotic conditions are in fact allowed for all $\mu>0$). With
these asymptotic conditions, the Hamiltonian is diagonalized by
the decomposition \beq\lab{mode}
    \Psic(\eta,\theta) = \int_{-\infty}^{\infty} \frc{d\nu}{\sqrt{2\pi}} c_\nu \U_\nu(\eta) e^{-\nu\theta} \;,\
    \Psic^\dag(\eta,\theta) = \int_{-\infty}^\infty \frc{d\nu}{\sqrt{2\pi}} c_\nu^\dag \U_\nu^\dag(\eta) e^{\nu\theta}
\eeq in terms of partial waves
\[
    \U_\nu = \mato{c} u_\nu\\v_\nu \matf,
\]
with \beqa\lab{unu}
    u_\nu &=& \frc{\Gamma(1+\mu)\Gamma\lt(\frc12+\mu-i\nu\rt)}{\Gamma(1+2\mu)\Gamma\lt(\frc12-i\nu\rt)}
        e^{i\nu\eta}(1-e^{2\eta})^{\mu} F(\mu,\frc12+\mu+i\nu;1+2\mu;1-e^{2\eta}) \\
\lab{vnu}
    v_\nu &=& - i \frc{\Gamma(1+\mu)\Gamma\lt(\frc12+\mu-i\nu\rt)}{\Gamma(1+2\mu)\Gamma\lt(\frc12-i\nu\rt)}
        e^{-i\nu\eta}(1-e^{2\eta})^{\mu} F(\mu,\frc12+\mu-i\nu;1+2\mu;1-e^{2\eta}),
\eeqa where $F(a,b;c;z)$ is Gauss's hypergeometric function on its
principal branch. In \eq{mode}, the operators $c_\nu,\,c_\nu^\dag$
satisfy the canonical anti-commutation relations
$$
    \{ c_\nu^\dag,c_{\nu'}\} = \delta(\nu-\nu').
$$
The angular Hilbert space ${\cal H}_A$ is the fermionic Fock space
over this algebra, with vacuum vector $|0\>_A$ defined by
$$
    c_\nu |0\>_A = 0 \quad (\nu > 0) ,\qquad c_\nu^\dag |0\>_A = 0 \quad (\nu < 0) .
$$
With an appropriate normal-ordering, the Hamiltonian takes the
form
\[
    H_A = \int_{0}^{\infty} d\nu \,\nu (c_\nu^\dag c_\nu + c_{-\nu} c^\dag_{-\nu}).
\]

We now consider the embedding $\Hperp \hookrightarrow {\cal H}_A
\otimes {\cal H}_A^*$ that will allow us to calculate form factors
of scaling fields
$\<vac|\Or_\a(0)|\om_1,\ldots,\om_n\>_{\ep_1,\ldots,\ep_n}$. The
embedding is described by identifying vectors in the Hilbert space
$\Hperp$ with endomorphisms on the angular Hilbert space ${\cal
H}_A$: \beq\lab{embed}
    |\om_1,\ldots,\om_n\>_{\ep_1,\ldots,\ep_n} \equiv a_{\ep_1}(\om_1) \cdots a_{\ep_n}(\om_n) e^{-\pi H_A},
\eeq where the operators $a_\ep(\om) \in \End({\cal H}_A)$ are to
be determined. Notice that the vacuum $|vac\>$ is identified with
$e^{-\pi H_A}$. The scalar product on $\Hperp$ is identified with
the canonical scalar product on the space $\End({\cal H}_A)$,
which coincides with the expression of correlation functions as
traces on ${\cal H}_A$: \beq\lab{form}
    \<u|v\> \equiv \frc{\Tr\lt( U^\dag V\rt)}{\Tr\lt(e^{-2\pi H_A}\rt)}\;
    \mbox{ if } |u\> \equiv U,\, |v\> \equiv V.
\eeq The representation of a field on $\Hperp$ is identified with
its representation on ${\cal H}_A$:
$$
    \piperp(\Or) |u\> \equiv \pi_A(\Or) U\; \mbox{ if } |u\> \equiv U,
$$
where $\pi_A$ is the representation map on the space ${\cal H}_A$.

The operators $a_\ep(\om)$ can be fixed by imposing two
conditions. First, the operators on ${\cal H}_A$ representing
fields at the center of the disk that are mutually local with the
fermion field must commute (if they are bosonic) or anti-commute
(if they are fermionic) with the operators $a_\ep(\om)$. It is
sufficient to impose this condition with the fermion operators
themselves: \beq\lab{loc}
    \{a_\ep(\om), \Psic(\eta\to-\infty)\} = \{a_\ep(\om), \Psic^\dag(\eta\to-\infty)\} = 0.
\eeq Second, the embedding \eq{embed} must reproduce the form
factors of fermion fields
$\<vac|\Psi_s(\xi_\rx,\xi_\ry)|\om\>_+=e^{-i\om \xi_\rx}
P_+(\om,\xi_\ry)$ and $\<vac|\Psi_s^\dag(\xi_\rx,\xi_\ry)|\om\>_-
= e^{-i\om \xi_\rx} P_-(\om,\xi_\ry)$ obtained from the partial
wave decomposition \eq{pwave}. As shown in Appendix A, these two
conditions are satisfied by the operators \beq\lab{Acv}
    a_+(\om) = \int_{-\infty}^{\infty} d\nu\, g(\nu;\om) \, c_\nu^\dag
    \,,\ \ \
    a_-(\om) = \int_{-\infty}^{\infty} d\nu\, g(\nu;\om) \, c_{-\nu},
\eeq
$$
    g(\nu;\om) =  \sqrt{\pi} 2^{-\mu} e^{i\frc{\pi}2 \lt(\mu+\frc12-i\frc{\om}2\rt)}
        \frc{e^{-\pi\nu} \G\lt(\frc12+\mu+i\nu\rt)}{
        \G\lt(1+\mu\rt) \G\lt(\frc12+i\nu\rt)}
        F\lt(\mu+\frc12+i\nu,\mu+\frc12-i\frc{\om}2;1+2\mu;2-i0\rt) .
$$
Here and below we use the notation
$$
    F(a,b;c;2\pm i0) = \lim_{\varepsilon\to0^+} F(a,b;c;2\pm i\varepsilon),
$$
where on the right hand side $F(a,b;c;z)$ is Gauss's
hypergeometric function on its principal branch. We believe that
the solution \eq{Acv} to the two conditions above is unique. One
can verify for instance that
$$
    \frc{\Tr\lt( e^{-2\pi H_A} \pi_A(\Psi_s(\xi_\rx,\xi_\ry)) a_+(\om)\rt)}{\Tr\lt(e^{-2\pi H_A}\rt)}
    = e^{-i\om \xi_\rx} P_+(\om,\xi_\ry);
$$
such expression can be calculated by using the cyclic properties
of the traces and the anti-commutation relations for the modes
$c_\nu, c^\dag_\nu$. As is shown in Appendix B.1, the operators
$a_\ep(\om)$ anti-commute among themselves: \beq\lab{Aacomm}
    \{a_{\ep_1}(\om_1), a_{\ep_2}(\om_2)\} = 0.
\eeq Also, since they are linear combinations of free modes,
traces of products of such operators satisfy Wick's theorem. In
particular we have the following contractions: \beqa &&
\frc{\Tr\lt( e^{-2\pi H_A} a_{\ep}^\dag(\om)
a_{\ep'}(\om')\rt)}{\Tr\lt( e^{-2\pi H_A}\rt)}
    = \frc1{\rho(\om)} \delta(\om-\om') \delta_{\ep,\ep'} \no\\
&&  \frc{\Tr\lt( e^{-2\pi H_A} a_{\ep}(\om)
a_{\ep'}(\om')\rt)}{\Tr\lt( e^{-2\pi H_A}\rt)}
    = 0
\eeqa Since these contractions are respectively ${\
}_\ep\<\om|\om'\>_{\ep'}$ and $\<vac|\om,\om'\>_{\ep,\ep'}$ and
since any form factor of fermion fields in $\Hperp$ can be
evaluated by Wick's theorem, the embedding \eq{embed} reproduces
all form factors of fermion fields.

From the expression \eq{Acv} one can verify the following
identity:
$$
    a^\dag_\ep(\om) = e^{\pi H_A} a_{-\ep}(\om) e^{-\pi H_A}.
$$
This is the analogue of crossing symmetry present in theories on
flat space. For the scaling fields $\Or_\a$, which are scalar and
have the property $\Or_\a^\dag = \Or_{-\a}$, this leads for
instance to \beq\lab{cs}
    (F_\a(\om_1,\ldots,\om_n)_{\ep_1,\ldots,\ep_n})^* = F_{-\a}(\om_n,\ldots,\om_1)_{-\ep_n,\ldots,-\ep_1}
\eeq using the notation \eq{Fff}.

From the representation \eq{scalingfieldPsi} of the scaling fields
on $\Hperp$, we have the following embedding:
$$
    \Or_\a(0)|u\> \equiv e^{2\pi i \a Q} U \; \mbox{ if } |u\>\equiv U,
$$
where $Q$ is the $U(1)$ charge in angular quantization:
\beq\lab{Q}
    Q = \int_{0}^{\infty} d\nu \,(c_\nu^\dag c_\nu - c_{-\nu} c^\dag_{-\nu}).
\eeq Using this embedding, form factors of the scaling fields
$\Or_\a$ are given heuristically by the following traces:
$$
    \<vac|\Or_\a(0) |\om_1,\ldots,\om_n\>_{\ep_1,\ldots,\ep_n}
        = \frc{\Tr\lt(e^{-2\pi H_A +2\pi i \a Q} a_{\ep_1}(\om_1) \cdots a_{\ep_n}(\om_n) \rt)}{
        \Tr\lt(e^{-2\pi H_A}\rt)}.
$$
Both traces in the ratio above are ill-defined and need some
regularization $\Tr\to \Tr_\varepsilon$. In the next section we
will regularize such traces by doing the angular quantization on a
Poincar\'e disk from which a small disk of radius $\varepsilon$
around the origin has been removed. As the regularization
parameter disappears $\varepsilon\to0$, the resulting ratio of
traces above is then singular and goes as $\varepsilon^{\a^2}$. We
can cancel out this singularity by considering normalized form
factors: \beq\lab{ffdef}
    \frc{\<vac|\Or_\a(0) |\om_1,\ldots,\om_n\>_{\ep_1,\ldots,\ep_n}}{\<vac|\Or_\a|vac\> }
        = \frc{\Tr\lt(e^{-2\pi H_A +2\pi i \a Q} a_{\ep_1}(\om_1) \cdots a_{\ep_n}(\om_n) \rt)}{
        \Tr\lt(e^{-2\pi H_A +2\pi i \a Q}\rt)}.
\eeq These can be calculated without explicit reference to a
regularization procedure, simply by using, as above, the cyclic
properties of the trace and the anti-commutation relations for the
modes $c_\nu$, $c_\nu^\dag$. The calculation of the one-point
function requires the use of an explicit regularization procedure:
\beq\lab{trace}
    \<\Or_\a\> = \<vac|\Or_\a|vac\> = \lim_{\varepsilon\to0} \varepsilon^{-\a^2} \frc{\Tr_\varepsilon\lt(e^{-2\pi H_A +2\pi i \a Q} \rt)}{\Tr_\varepsilon\lt(e^{-2\pi H_A}\rt)},
\eeq and will be done in the next section.

The traces \eq{ffdef} are calculated in Appendix B. In particular,
the ``two-particle'' form factors are given in \eq{ffapp}. Other
``multi-particle'' form factors of scaling fields $\Or_\a(0)$ can
be constructed from these ``two-particle'' form factors by Wick's
theorem, as in \eq{mpff}. In Appendix C, we verify that the
expression \eq{ffapp} for the ``two-particle'' form factors
specializes to the known expression for form factors in the
flat-space limit.

\sect{One-point function}

We now calculate the one-point function $\<O_\a\>$. This is a
simple generalization of the calculation done in \cite{AlZ,LZ} for
similar vacuum expectation values in flat space.

We regularize the traces $\Tr\to\Tr_\varepsilon$ by cutting a
small disk of radius $\varepsilon$ around the origin and
considering the angular quantization of the theory \eq{theory} on
the resulting annulus with the boundary conditions
\beq\lab{boundcond}
    [\PsicR(\eta) - \PsicL(\eta)]_{\eta=\ln\,\tanh\lt(\frc{\varepsilon}{2R}\rt)} =
    [\PsicdR(\eta) - \PsicdL(\eta)]_{\eta=\ln\,\tanh\lt(\frc{\varepsilon}{2R}\rt)} = 0.
\eeq Then the one-point function $\<\Or_\a\>$ can be expressed as
\eq{trace}, where the $U(1)$ charge $Q$ in the regularized theory
is
$$
    Q = \int_{\ln\,\tanh\lt(\frc{\varepsilon}{2R}\rt)}^{0} d\eta  : \Psic^\dag \Psic :,
$$
which specializes to \eq{Q} in the limit $\varepsilon\to0$. It was
shown in \cite{AlZ,LZ} that for the theory on flat space, the
definition \eq{trace} with the boundary conditions \eq{boundcond}
are in accordance with the conformal normalization \eq{norm}.
Since the leading behavior at short distances of the two-point
function is not affected by the curvature, the expression
\eq{trace} with the boundary conditions \eq{boundcond} lead to the
same conformal normalization \eq{norm} for the theory on the
Poincar\'e disk.

In order to evaluate the limit $\varepsilon\to0$ in the expression
\eq{trace}, we need the density of angular quantization states
$\d_\nu\ln(S(\nu))$, where the S-matrix $S(\nu)$ is associated
with the scattering off the ``mass barrier'' described in the
previous section:
\[
    \lt. \mato{c} u_\nu\\v_\nu \matf \rt|_{\eta\to-\infty} = \mato{c} e^{i\nu\eta} \\ -ie^{-i\nu\eta} S(\nu) (2R)^{-2i\nu} \matf .
\]
Here $u_\nu$ and $v_\nu$ are the partial waves \eq{unu} and
\eq{vnu}. The S-matrix is given by: \beq\lab{Smatrix}
    S(\nu) = (2R)^{2i\nu} \frc{\Gamma(1/2+i\nu)\Gamma(1/2-i\nu+\mu)}{\Gamma(1/2-i\nu)\Gamma(1/2+i\nu+\mu)} .
\eeq Then one finds \beq \lab{1ptfct}
    \<\Or_\a\> = \exp\lt[\int_0^\infty \frc{d\nu}{2\pi i}
        \ln\lt( \frc{\lt(1+e^{-2\pi\nu + 2\pi i \a}\rt) \lt(1+e^{-2\pi\nu - 2\pi i \a}\rt)}{ \lt(1+e^{-2\pi\nu}\rt)^2 } \rt)
        \d_\nu \ln S(\nu) \rt].
\eeq The result of the integration using \eq{Smatrix} can be
expressed in various ways: \beqa
    \<\Or_\a\> &=& (2R)^{-\a^2} \exp\lt[ \int_0^\infty \frc{dt}{t} (1-e^{-2\mu t}) \frc{\sinh^2(\a t)}{\sinh^2(t)}\rt] \no\\
\lab{VEV}   &=& (2R)^{-\a^2} \prod_{n=1}^{\infty} \lt( \frc{1-\frc{\a^2}{(\mu+n)^2}}{1-\frc{\a^2}{n^2}}\rt)^n \\
        &=& (2R)^{-\a^2} \frc{G(1+\mu-\a) G(1+\mu+\a)}{G(1+\mu)^2 G(1-\a) G(1+\a)} \no,
\eeqa where $G(z)$ is Barnes' $G$-function, characterized mainly
by the properties $G(z+1)=\G(z)G(z)$ and $G(1)=1$ (cf.
\cite{SrivastavaChoi}).

\sect{Long distance expansion of the two-point function}

As argued in Appendix B.1, the ``two-particle'' form factors
\eq{ffapp}, like the partial waves \eq{pwaveexpr}, are entire
functions of the spectral parameters $\om_1$ and $\om_2$ (hence
all form factors of scaling fields $\Or_\a$ are entire functions
of their spectral parameters). Moreover, they have the following
behavior as the real part of $\om_1$ or $\om_2$ goes to positive
or negative infinity: \beqa
&&  \sqrt{\rho(\om_1)} F_\a(\om_1,\om_2)_{+,-} \sim |\om_1|^{-1\pm\a} \; \mbox{ as } \Re e(\om_1)\to \pm\infty\no\\
&&  \sqrt{\rho(\om_2)} F_\a(\om_1,\om_2)_{+,-} \sim
|\om_2|^{-1\mp\a} \; \mbox{ as } \Re e(\om_2)\to \pm\infty\no
\eeqa up to proportionality factors, and other ``multi-particle''
form factors have similar behaviors. Hence for $-1<\a-\a'<1$, the
integrals over $\om_j$'s in \eq{genexp} are absolutely convergent.
They can be evaluated by deforming the contours in their lower
half planes and summing over the residues at the poles of the
measure $\rho(\om_j)$. The measure $\rho(\om)$ has poles on the
imaginary axis; in the lower half plane, they are at positions
$\om = -i(1+2\mu+2k)$ for $k\in \N$.

It is instructive to interpret the poles in the measure
$\rho(\om)$, and the evaluation of the integrals in \eq{genexp} by
contour deformation as described above, in terms of a different
quantization scheme. Let us summarize our construction. We choose
an isometry $\iK$ \eq{isoK}, subgroup of the full isometry group
$SU(1,1)$, and we quantize the theory \eq{theory} on orbits of
this isometry. That is, the generator $\genK$ of the subgroup
$\iK$ generates ``space'' translations. A basis for the
corresponding Hilbert space $\Hperp$ is taken as a set of states
diagonalizing $\genK$. They are normalized so that all form
factors of local fields are entire functions of its eigenvalues
$\om$. With this normalization, the resolution of the identity on
$\Hperp$ in terms of this basis involves a measure $\rho(\om)$
\eq{meas} with a specific analytical structure, in particular with
singularities on the imaginary axis. Now, these singularities
should give information about the spectrum in a quantization
scheme where $\genK$ is taken as the Hamiltonian, that is, as the
generator of ``time'' translations.

This last assertion is the analogue of what is well known to
happen for instance in a free massive theory on flat space: the
singularity structure of the invariant measure $\rho(p) =
(m^2+p^2)^{-\frc12}$ as function of the momentum $p$ gives the
energy spectrum. Momentum operator and energy operator (or
Hamiltonian) can be seen as representations of the same
translation generator in two different quantization scheme, one
where this translation is along the ``space'' direction, the other
where it is along the ``time'' direction.

In a free theory on flat space, the invariant measure exhibits in
particular a branch cut starting at $p=-im$ and going to
$p\to-i\infty$, corresponding to the continuous spectrum (from $m$
to infinity) of the energy operator. In our free theory on the
Poincar\'e disk, the position of the poles of the measure
\eq{meas} on the imaginary axis are interpreted as the discrete
eigenvalues of the generator $\genK$ in the scheme where it is
taken as the Hamiltonian. It is a simple matter to repeat the
canonical quantization procedure of section 3 for this
quantization scheme. One indeed finds a discrete set of
eigenstates
$|k_1,\ldots,k_n\>_{\ep_1,\ldots,\ep_n},\;k_j\in\N,\;n=0,1,2,\ldots$
with eigenvalues $\l_1+\cdots+\l_n$, where
$$
    \l_j = 1+2\mu + 2k_j.
$$
Since these states diagonalize a generator of time translation,
they can be more naturally interpreted as multi-particle states.
Matrix elements of the operator $\Or_\a$ between the vacuum $|0\>$
and multi-particle states in this scheme can be found from similar
matrix elements in the Hilbert space $\Hperp$ by the following
identification:
$$
    \<0|\Or_\a|k_1,\ldots,k_n\>_{\ep_1,\ldots,\ep_n} =
    \<\Or_\a\>\;
        \prod_{j=1}^n \lt( i^{k_j} \sqrt{\frc{2\G(1+2\mu+k_j)}{k_j!}}
        \frc{1}{\G\lt(\frc12+\mu\rt)}\rt)
        F_\a(-i\l_1,\ldots,-i\l_n)_{\ep_1,\ldots,\ep_n}
$$
and ${\ }_{\ep_1,\ldots,\ep_n}\<k_n,\ldots,k_1|\Or_\a|0\> =
\<0|\Or_{-\a}|k_1,\ldots,k_n\>_{\ep_1,\ldots,\ep_n}^*$. The
resolution of the identity is given by the sum
$$
    1 = \sum_{n=0}^{\infty} \frc1{n!} \sum_{\ep_1,\ldots,\ep_n}
    \sum_{k_1,\ldots,k_n \ge 0}
    |k_1,\ldots,k_n\>_{\ep_1,\ldots,\ep_n}
    {\ }_{\ep_1,\ldots,\ep_n}\<k_1,\ldots,k_n|,
$$
which provides a long distance expansion for the two-point
function: \beqa
    \<\Or_\a(x)\Or_{\a'}(y)\> &=& \sum_{n=0}^{\infty} \frc1{n!} \sum_{\ep_1,\ldots,\ep_n}
    \sum_{k_1,\ldots,k_n=0}^{\infty} \;\times  \no \\
&&\times\;  \<0|\Or_\a(0)|k_1,\ldots,k_n\>_{\ep_1,\ldots,\ep_n}
    {\ }_{\ep_n,\ldots,\ep_1}\<k_n,\ldots,k_1|\Or_{\a'}(0)|0\>
    e^{-(\l_1+\cdots+\l_n) \frc{\dist(x,y)}{2R}} \no.
\eeqa This expansion coincides with the sum of residues coming
from the evaluation of the integrals in \eq{genexp} by contour
deformation. Using crossing symmetry \eq{cs} and summing over the
$U(1)$ charges, this can be written \beq\lab{genexpHp}
    \<\Or_\a(x)\Or_{\a'}(y)\> = \<\Or_\a\>\<\Or_{\a'}\> \; \sum_{N=0}^{\infty} \frc1{(N!)^2}
    \sum_{k_1,\ldots,k_{2N}=0}^{\infty}
    f_\a(k_1,\ldots,k_{2N}) f_{\a'}(k_{2N},\ldots,k_1)
    e^{-(\l_1+\cdots+\l_{2N}) \frc{\dist(x,y)}{2R}},
\eeq where \beq\lab{f}
    \<\Or_\a\>\, f_\a(k_1,\ldots,k_{2N}) = \<0|\Or_\a(0)|k_1,\ldots,k_{2N}\>_{\underbrace{+,+,\ldots}_{N},\underbrace{-,-,\ldots}_{N}}.
\eeq The functions $f_\a(k_1,\ldots,k_{2N})$ are evaluated in
closed form in Appendices B.2, B.3. The first few terms of
\eq{genexpHp} give the following expansion for the two-point
function: \beqa\lab{ffexp} &&  F(t) = 1 - 4^{2\mu+1}
\frc{\sin(\pi\a) \sin(\pi\a') \G(1+\mu+\a) \G(1+\mu-\a)
    \G(1+\mu+\a') \G(1+\mu-\a') }{
    \pi^2 \G(2+2\mu)^2} e^{-(4\mu+2) t} \times \no\\
&&  \ \ \ \ \ \lt( 1 - 2\a\a' \frc{2\mu+1}{(\mu+1)^2} e^{-2t} +
O(e^{-4t}) \rt) \;
    + \; O(e^{-(8\mu+4)t}),
\eeqa where the function $F(t)$ was defined in
\eq{leadinglongdist}.

It is interesting to notice that in the massless limit $\mu\to0$,
the two-point function still decreases exponentially at long
distance. This is a signal of the infrared regulator properties of
the negative curvature of the Poincar\'e disk \cite{CW}.

\sect{Two-point function from Painlev\'e VI transcendents}

From \cite{PBT}, the two-point function can be described in terms
of a Painlev\'e VI transcendent: \beq\lab{diffeq}
    F(t) = \exp\lt[ -\int_{\tanh^2t}^1 ds \frc{d}{ds} \ln\tau(s) \rt],
\eeq where $\frc{d}{ds} \ln\tau(s)$ is given in \eq{tau} and the
Painlev\'e transcendent $w(s)$ satisfies \eq{painleveintro} with
$\ga=0$.

As said in the introduction, the integration constants specifying
the Painlev\'e transcendent can be taken to be the exponents in
the behaviors of $w$ at $s\sim 0$ (short distance) and at $s\sim
1$ (long distance), given, respectively, in \eq{ws0} and \eq{ws1}
(such a solution exists). The proportionality constants $B$ and
$A$ are then unambiguous. In particular, using results of Jimbo
\cite{J} there is an explicit expression for the constant $B$
involved in the short distance behavior of $w$: \beq\lab{B}
    B = \mu\frc{ \G(\a)\,\G(\a') \,\G(1-\a-\a')^2\, \G(\a+\a'+\mu)}{
        \G(1-\a)\,\G(1-\a')\, \G(\a+\a')^2\, \G(1-\a-\a'+\mu)}.
\eeq From this, one can obtain the full short distance expansion
of $w$, valid for $0<\a+\a'<1$:
$$
    w = Bs^{\a+\a'} \sum_{p,q=0}^\infty C_{p,q}\; s^{p (\a+\a') + q(1-\a-\a')}
$$
where $C_{0,0}=1$. Once $B$ is fixed, the other coefficients
$C_{p,q}$ are uniquely determined by the differential equation
\eq{painleveintro}.

The leading long distance asymptotics \eq{ws1} of this Painlev\'e
transcendent is fixed by specifying the proportionality constant
$A$. This constant can be obtained by comparing our form factor
result \eq{ffexp} with the expansion \eq{Flarged}; this leads to
the value \eq{A}. Since we must have $0<w<1$ for all $0<s< 1$, the
asymptotics \eq{ws1} is valid only for $A>0$, and the asymptotics
\eq{ws0} only for $B>0$. This imposes $\a\a'>0$. From the behavior
\eq{ws1} and using the differential equation, one can obtain the
full long distance expansion of $w$: \beq\lab{wlarged}
    1-w = A(1-s)^{1+2\mu}\sum_{p,q=0}^{\infty} D_{p,q} \; (1-s)^{p(1+2\mu) + q}
\eeq with $D_{0,0}=1$. For instance, the coefficients $D_{0,q}$
are given by \beq\lab{D0q}
    \sum_{q=0}^{\infty} D_{0,q}\; (1-s)^q = s^{\a-\a'}F(1-\a'+\mu,\a+\mu;1+2\mu;1-s)^2.
\eeq The long distance expansion \eq{wlarged} can also be obtained
from the full form factors expansion \eq{genexpHp}. For instance,
one can verify that the coefficient $D_{0,1}$ in \eq{D0q} is
consistent with the coefficient of $e^{-2t}$ in the parenthesis in
the form factor expansion \eq{ffexp}.

Solving numerically the differential equation by using the
appropriate initial conditions, it is possible to verify the
consistency of our result \eq{A} for the constant involved in the
long distance asymptotics of $w$, our result \eq{VEV} for the
one-point function, and the value \eq{B} for the constant involved
in the short distance asymptotics of $w$. We used as initial
condition the long distance asymptotics \eq{wlarged} with non-zero
coefficients $D_{0,q}$ given in \eq{D0q} and with the value \eq{A}
for the normalization constant. We numerically verified that the
behavior \eq{ws0} with the constant \eq{B} is recovered, and that
the equation \eq{norm} is satisfied to a high accuracy.

\sect{Conclusion}

We have fully characterized the two-point function
$\<\Or_\a(x)\Or_{\a'}(y)\>$ in the Dirac theory on the Poincar\'e
disk, with fermion mass $m$ and Gaussian curvature $-\frc1{R^2}$,
in the region $0<\a+\a'<1,\; \a\a'>0$ and $mR>\frc12$. A
comparison of our long distance expansion with the expansion
obtained from the Painlev\'e VI differential equation strongly
suggests both that our ``form factor'' expansion is correct and
that the Painlev\'e VI differential equation indeed describes the
two-point function of the scaling fields $\Or_\a$ in the region of
parameters above. Notice that our results and the description
above in terms of the differential equation are valid in the
region $0<mR<\frc12$ as well; however, they do not describe all
the possible behaviors in this region.

It would be very interesting to extend this description to a
larger region of $\a,\,\a'$, and to fully analyze the region
$0<mR<\frc12$ and the analytical continuation to negative values
of $mR$. Probably one of the most interesting eventual application
of this description is to the analysis of an appropriate scaling
limit of a ``near-critical'' lattice Ising model on the Poincar\'e
disk. Two-point functions of flat-space Ising spin and disorder
variables in the scaling limit (described by two-point functions
of spin and disorder fields in the free Majorana field theory) are
simply related to two-point functions of the scaling fields
$\Or_{\pm\frc12}$ in the free Dirac theory on flat space \cite{ZI,
ST, WMTB}. If similar relations exist between the Ising spin and
disorder variables on the Poincar\'e disk and the fields
$\Or_{\pm\frc12}$ in the free Dirac theory on the Poincar\'e disk,
the study of the latter can give information about the effect of
the Poincar\'e geometry on the statistical properties of the Ising
model. For instance, the one-point function \eq{VEV} for
$\a=\frc12$ should be proportional to the square of the Ising
magnetization near ``criticality''. In the same spirit, from the
point of view of the Ising model on the Poincar\'e disk, the
region $mR<\frc12$, and in particular the analytical continuation
to negative values of $mR$, should show the most interesting
features, especially in view of the arguments of \cite{CW}. Some
work has already been done in relating the scaling fields
$\Or_{\pm\frc12}$ in the Dirac theory on the Poincar\'e disk to
spin and disorder fields in the Majorana theory on the Poincar\'e
disk \cite{D}. We hope to present results of a more complete
investigation in a future publication \cite{DF}.

\

\bc \bf Acknowledgment \ec

I am very grateful to S. Lukyanov for suggesting the research
project to me and for numerous discussions. I have also extremely
benefited from discussions with P. Fonseca, A. B. Zamolodchikov
and G. Moore, and comments from C. A. Tracy. This research was
supported in part by an NSERC Postgraduate Scholarship.

\appendix

\sect{Construction of the embedding}

Here we describe a method for obtaining the expression \eq{Acv}
for the operators $a_\ep(\om)$. We first solve the locality
condition \eq{loc} for a set of operators $Z_\ep(r)$ depending on
a complex parameter $r$:
$$
    \{Z_\ep(r), \Psic(\eta\to-\infty)\} = \{Z_\ep(r), \Psic^\dag(\eta\to-\infty)\} = 0,
$$
with the following ansatz:
$$
    Z_+(r) = \int_{-\infty}^\infty d\nu \, f(\nu) \, c^\dag_\nu\,r^{- i \nu}\;,\
    Z_-(r) = \int_{-\infty}^\infty d\nu \, f(\nu) \, c_{-\nu}\,r^{- i \nu}.
$$
This leads to the equations:
$$
    \lim_{\eta\to-\infty} \int_{-\infty}^\infty d\nu\, e^{i\nu\eta}f(\nu) r^{-i\nu} = 0,\
    \lim_{\eta\to-\infty} \int_{-\infty}^\infty d\nu\, S(\nu) (2R)^{-2i\nu} e^{-i\nu\eta} f(\nu) r^{-i\nu} = 0,
$$
where $S(\nu)$ is given in \eq{Smatrix}. The first equation is
satisfied if $f(\nu)$ is analytical in the lower half $\nu$-plane
and increases at most exponentially as $\Im m(\nu)\to-\infty$, and
the second equation is satisfied if $S(\nu) f(\nu)$ is analytical
in the upper-half $\nu$-plane and increases at most exponentially
as $\Im m(\nu)\to\infty$. Indeed, under such conditions it is
possible, for $\eta$ negative and large enough, to send the
contour of integration to $\Im m(\nu) \to -\infty$ in the integral
$\int_{-\infty}^\infty d\nu\, e^{i\nu\eta}f(\nu) r^{-i\nu}$,
giving zero contribution, and similarly for the integral
$\int_{-\infty}^\infty d\nu\, S(\nu) (2R)^{-2i\nu} e^{-i\nu\eta}
f(\nu) r^{-i\nu}$ by sending the contour of integration to $\Im
m(\nu) \to \infty$. This set of conditions on $f(\nu)$ forms a
simple Riemann-Hilbert problem, a solution of which is: \beq
    f(\nu) = \sqrt{2\pi} \frc{\G\lt(\frc12+\mu+i\nu\rt)}{\G\lt(\frc12 + \mu\rt) \G\lt(\frc12+i\nu\rt)}.
\lab{fnu} \eeq

The operators $a_\ep(\om)$ can then be formed by taking
appropriate linear combinations of $Z_\ep(r)$. These linear
combinations can be obtained by requiring that the states
$|\om\>_\ep \equiv a_\ep(\om) e^{-\pi H_A}$ diagonalize the
subgroup $\iK$ and that they be correctly normalized. First
consider states $|r\>_\ep \in \Hperp$ embedded in ${\cal H}_A
\otimes {\cal H}_A^*$ by the identification $|r\>_\ep \equiv
Z_\ep(r)e^{-\pi H_A}$. Using this embedding we can calculate the
following matrix elements: \beqa
    \<vac| \piperp(\Psi_R(z,\b{z})) |r\>_+ &=& -i \frc{\G(1+\mu)}{\G\lt(\frc12+\mu\rt)}
        (1-z\b{z})^\mu r^{-\frc12} (1-\b{z} r)^{-\mu} (1-zr^{-1})^{-\mu-1} \no\\
    \<vac| \piperp(\Psi_L(z,\b{z})) |r\>_+ &=& \frc{\G(1+\mu)}{\G\lt(\frc12+\mu\rt)}
        (1-z\b{z})^\mu r^{\frc12} (1-z r^{-1})^{-\mu} (1-\b{z}r)^{-\mu-1}.
\lab{PsiZ}\eeqa From this and from the transformation properties
of the fermion fields one can infer the transformation properties
of the states $|r\>_\ep$ under the isometry group $SU(1,1)$:
$$
    \h{g} \lt| \frc{ ar + \b{b}}{ br + \b{a}} \rt\>_\ep = H_{\frc{\mu}2+\frc14,g}(r) \b{H}_{\frc{\mu}2+\frc14,g}(r^{-1}) |r\>_\ep,
$$
where the functions $H_{s,g}$ and $\b{H}_{s,g}$ are automorphic
factors:
$$
    H_{s,g}(z) = (bz+\b{a})^{2s} \;,\  \b{H}_{s,g}(\b{z}) = (\b{b} \b{z} + a)^{2s}.
$$
Diagonalizing this action for the subgroup $\iK$, we obtain the
operators $a_\ep(\om)$: \beq\lab{aint}
    a_\ep(\om) = \frc{2^{-\mu-\frc12} e^{-i\frc{\pi}2 \lt(\mu+\frc12 + i\frc{\om}2\rt) } \G\lt(\frc12+\mu\rt)^2}{
    \sqrt{\pi}\G\lt(\frc12 + \mu + i\frc{\om}2\rt) \G\lt(\frc12 + \mu - i\frc{\om}2\rt)}
    \int_{-1}^1 dr\,(1-r)^{\mu-\frc12+i\frc{\omega}2} (1+r)^{\mu-\frc12-i\frc{\om}2}r^{-\mu-\frc12} Z_\ep(r),
\eeq where the integral is performed in the region
$-\pi<\arg(r)<0$ (other integration contours are possible but
equivalent). The normalization was fixed by comparing with the
partial wave \eq{pwaveexpr}. The expression \eq{aint} for
$a_\ep(\om)$ is equivalent to \eq{Acv}.

\sect{Construction of form factors} We study the normalized form
factors of scaling fields \eq{ffdef}:
$$
    F_\a(\om_1,\ldots,\om_n)_{\ep_1,\ldots \ep_n}
    = \<\<a_{\ep_1}(\om_1) \cdots a_{\ep_n}(\om_n)\>\>_\a,
$$
where we use the notation \beq\lab{normcorr}
    \<\< \cdots \>\>_\a =
        \frc{\Tr\lt(e^{-2\pi K + 2\pi i\a Q} \cdots \rt)}{\Tr\lt( e^{-2\pi K + 2\pi i\a Q}\rt)}.
\eeq We will assume that the spectral parameters $\om_j$ are real,
except when explicitly stated.

\ssect{Construction of ``two-particle'' form factors}

From the cyclic properties of the trace and the anti-commutation
relations for the angular quantization modes, we have
\[
    \<\<c_\nu^\dag c_{\nu'}\>\>_\a = \frc{\delta(\nu-\nu')}{1+e^{2\pi\nu-2\pi i\a}}.
\]
From the definition \eq{Acv}, the trace \eq{ffdef} is then
expressed as an integral: \beqa &&  \<\<a_+(\om_1)a_-(\om_2)
\>\>_\a =
    \frc{4^{-\mu} e^{(\om_2-\om_1)\frc\pi4}}{\G(1+\mu)^2} \int_{-\infty}^\infty d\nu\,
    \frc{\G\lt(\mu+\frc12+i\nu\rt) \G\lt(\mu+\frc12-i\nu\rt) \cosh(\pi\nu)}{1+e^{2\pi\nu-2\pi i \a}}
    \times \no\\
&&  \qquad F\lt( \mu + \frc12 -
i\nu,\mu+\frc12-i\frc{\om_1}2;1+2\mu;2+i0\rt)
    F\lt( \mu + \frc12 - i\nu,\mu+\frc12-i\frc{\om_2}2;1+2\mu;2-i0\rt). \no
\eeqa The contour of integration goes between the poles at
$\nu=i\a-i/2$ and $\nu=i\a+i/2$.

The integrand is proportional to
$e^{-\pi\nu}(-\nu)^{-1+i(\om_1-\om_2)/2}$ when $\nu\to-\infty$,
and to $e^{-\pi\nu}(\nu)^{-1-i(\om_1-\om_2)/2}$ when
$\nu\to\infty$. This can be obtained from the asymptotics
$$
    F(a,b;c;2\pm i0) = \frc{\G(c)}{\G(b)} (2a)^{b-c} e^{\pm i\pi(a+b-c)}\lt(
        1+O(a^{-1}) \rt) +
         \frc{\G(c)}{\G(c-b)} (2a)^{-b} e^{\pm i\pi b}\lt(
        1+O(a^{-1}) \rt) ,
$$
valid for $|a|\to \infty,\; |\arg(a)|<\pi$. The integral can be
regularized by multiplying the integrand by a factor $e^{p\nu}$
with some complex parameter $p$; the integral is then convergent
for $\Re e(p) = \pi,\, \Im m(p) \neq0$. It can be evaluated by the
method of residues, closing the contour for instance in the upper
half plane if $\Im m(p) >0$. The ``two-particle'' form factor is
the analytical continuation to $p=0$ of the resulting expression.
Contributions of poles in the upper half plane give
\beq\lab{ffapp0}
    \<\<a_+(\om_1)a_-(\om_2) \>\>_\a =
    \frc{4^{-\mu} i\pi e^{(\om_2-\om_1)\frc\pi4}}{
        \sin(\pi(\mu-\a)) \G(1+\mu)^2}
    \lt[ G_{\a}(\om_1,\om_2) - e^{-i\pi(\mu-\a)} G_{\mu}(\om_1,\om_2) \rt],
\eeq where $G_{\a}(\om_1,\om_2)$ is the analytical continuation to
$p=0$ of the following series: \beqa\lab{G} &&
G_\a^{(p)}(\om_1,\om_2) = \sum_{n=0}^\infty \lt[ \sin(\pi\a)
        \frc{\G(1+\mu+\a+n)}{\G(1-\mu+\a+n)} e^{ip(\frc12+\a+n)} \rt. \\
&&\times\;  \lt.
F\lt(\mu+1+\a+n,\mu+\frc12-i\frc{\om_1}2;2\mu+1;2+i0\rt)
        F\lt(\mu+1+\a+n,\mu+\frc12-i\frc{\om_2}2;2\mu+1;2-i0\rt) \rt]
        \no
\eeqa and $G_\mu(\om_1,\om_2)$ is the analytical continuation to
$p=0$ of the series above with $\a$ replaced by $\mu$. Note that
the function $G_\mu(\om_1,\om_2)$ is shown below to be identically
zero.

The series $G_\a^{(p)}(\om_1,\om_2)$ is convergent in the upper
half $p$-plane $\Im m(p)>0$ as well as on $\Im m(p)=0,\; \Re
e(p)\neq\{0,\pi,-\pi\}$. The analytical continuation to $p=0$ can
be done via a ``zeta-regularization''. More precisely, we
subtract, inside the summation symbol in \eq{G}, the leading large
$n$ asymptotics of the summand: \beqa
&&  \G(2\mu+1)^2 \sin(\pi\a) 4^{-\mu} (2n)^{-1} e^{\pi\Delta}e^{ip(\frc12+\a+n)} \times \no\\
&&  \lt[ \frc{(2n)^{-i\om}}{\G\lt(\mu+\frc12-i\frc{\om_1}2\rt)
\G\lt(\mu+\frc12 - i\frc{\om_2}2\rt)}
         + \rt.
    \frc{(2n)^{i\om}}{\G\lt(\mu+\frc12+i\frc{\om_1}2\rt) \G\lt(\mu+\frc12 + i\frc{\om_2}2\rt)}
         + \no \\
&&
\frc{e^{i\pi(\a-\mu+n)}(2n)^{-i\Delta}}{\G\lt(\mu+\frc12-i\frc{\om_1}2\rt)
\G\lt(\mu+\frc12 + i\frc{\om_2}2\rt)}
          +
    \lt. \frc{e^{-i\pi(\a-\mu+n)}(2n)^{i\Delta}}{\G\lt(\mu+\frc12+i\frc{\om_1}2\rt) \G\lt(\mu+\frc12 - i\frc{\om_2}2\rt)}
          \rt] \no,
\eeqa where $\om = (\om_1+\om_2)/2$ and $\Delta =
(\om_1-\om_2)/2$. The resulting series is convergent at $p=0$. We
then add to this series at $p=0$ the following quantity: \beqa
&&  \G(2\mu+1)^2 \sin(\pi\a) 4^{-\mu} e^{\pi\Delta} \times \no\\
&&  \lt[ \frc{1}{\G\lt(\mu+\frc12-i\frc{\om_1}2\rt)
\G\lt(\mu+\frc12 - i\frc{\om_2}2\rt)}
        \frc{\zeta(1+i\om)}{2^{1+i\om}} + \rt.
    \frc{1}{\G\lt(\mu+\frc12+i\frc{\om_1}2\rt) \G\lt(\mu+\frc12 + i\frc{\om_2}2\rt)}
        \frc{\zeta(1-i\om)}{2^{1-i\om}} +  \no \\
&&  \frc{e^{i\pi(\a-\mu)}(2^{-i\Delta}-1)
}{\G\lt(\mu+\frc12-i\frc{\om_1}2\rt) \G\lt(\mu+\frc12 +
i\frc{\om_2}2\rt)}
        \frc{\zeta(1+i\Delta)}{2^{1+i\Delta}} +
    \lt. \frc{e^{-i\pi(\a-\mu)}(2^{i\Delta}-1) }{\G\lt(\mu+\frc12+i\frc{\om_1}2\rt) \G\lt(\mu+\frc12 - i\frc{\om_2}2\rt)}
        \frc{\zeta(1-i\Delta)}{2^{1-i\Delta}} \rt]  \no,
\eeqa where $\zeta(z)$ is Riemann's zeta function. The result is
$G_\a(\om_1,\om_2)$. Notice that the function $G_\a(\om_1,\om_2)$
is real for real $\om_1,\om_2$.

The resulting expression for $G_\a(\om_1,\om_2)$ defines a
function of $\om_1$ and $\om_2$ analytical in the region $|\Im m
(\om_1+\om_2)|<2$, $|\Im m (\om_1-\om_2)| < 2$. By repeating the
procedure above for the full large $n$ asymptotics of the summand
of $G_\a^{(p)}(\om_1,\om_2)$, one can see that the function
$G_\a(\om_1,\om_2)$ thus defined has no singularity in the finite
complex $\om_1$- and $\om_2$-planes: the function
$G_\a(\om_1,\om_2)$ is an entire function of $\om_1$ and $\om_2$.

The function $G_\mu(\om_1,\om_2)$ can now be evaluated in the
following way. From \eq{ffapp0}, we have
\[
    \<\<a_+(\om_1)a_-(\om_2) \>\>_0 =
    -\frc{4^{-\mu} i\pi e^{(\om_2-\om_1)\frc\pi4}}{
        \sin(\pi\mu) \G(1+\mu)^2}
    e^{-i\pi\mu} G_{\mu}(\om_1,\om_2).
\]
By covariance under the subgroup $\iK$ we find
\[
    (\om_1+\om_2) G_{\mu}(\om_1,\om_2) = 0.
\]
Since the function $G_{\mu}(\om_1,\om_2)$ is entire, this implies
that it must be identically zero. Notice that a similar
calculation leads to $\<\<a_-(\om_1)a_+(\om_2)\>\>_0=0$. Since the
commutator $\{a_-(\om_1),a_+(\om_2)\}$ is a $c$-number, an
immediate consequence is the anti-commutation property
\eq{Aacomm}.

Hence we finally have \beq
    F_\a(\om_1,\om_2)_{+,-} =
    \frc{4^{-\mu} i\pi e^{(\om_2-\om_1)\frc\pi4}}{
        \sin(\pi(\mu-\a)) \G(1+\mu)^2} G_{\a}(\om_1,\om_2).
\lab{ffapp} \eeq

\ssect{Two-particle form factors in the discrete basis}

It is possible to evaluate the function $G_\a(\om_1,\om_2)$ for
some purely imaginary values of $\om_1$ and $\om_2$. We will
evaluate it for the values $\om_1 = -i(1+2\mu+2k_1)$ and $\om_2 =
-i(1+2\mu+2k_2)$ for integers $k_1\ge0$ and $k_2\ge 0$. This can
be done by using the analytical continuation described above.
Equivalently, it can be done by simply replacing these values of
$\om_1$ and $\om_2$ in the expression \eq{G}, evaluating the
resulting series at $p=0$ in a region of $\mu$ where it is
convergent and analytically continuing the result in $\mu$. With
$$
    G_{\a;\,k_1,k_2} \equiv G_\a(-i(1+2\mu+2k_1),-i(1+2\mu+2k_2)),
$$
this gives \beq
    G_{\a;\,k_1,k_2}  =
    \sin(\pi\a) \sum_{m_1=0}^{k_1} \sum_{m_2=0}^{k_2}
    \frc{(-k_1)_{m_1} (-k_2)_{m_2} 2^{m_1+m_2}}{
    (2\mu+1)_{m_1} (2\mu+1)_{m_2} m_1! m_2!} H_{\a;\,m_1,m_2}
\eeq where
$$
    H_{\a;\,m_1,m_2} = \frc{\G(1+\mu+\a+m_1) \G(1+\mu+\a+m_2)}{\G(1-\mu+\a) \G(1+\mu+\a)}
    {\ }_3F_2(1,1+\mu+\a+m_1,1+\mu+\a+m_2;1+\mu+\a,1-\mu+\a;1).
$$
The ${\ }_3F_2$ hypergeometric function above can be evaluated in
closed form, for any given integer $m_1$ and $m_2$, in terms of
Gamma functions and rational functions of $\mu$ and $\a$. The
two-particle form factors $\eq{f}$ in the discrete basis are
expressed in terms of $G_{\a;k_1,k_2}$: \beq
    f_\a(k_1,k_2) =
    2^{2\mu+1}i(-1)^{k_1} \sqrt{\frc{\G(1+2\mu+k_1)\G(1+2\mu+k_2)}{k_1!k_2!}}
    \frc{G_{\a;\,k_1,k_2}}{\G(1+2\mu)^2 \sin(\pi(\mu-\a)) } .
\eeq

\ssect{``Multi-particle'' form factors}

The ``multi-particle'' form factors of scaling fields can be
evaluated by Wick's theorem in terms of the two-particle form
factors. We have \beq
    F_\a(\om_1,\ldots,\om_{n})_{\ep_1,\ldots,\ep_n}
    = \sum_{j=2}^n (-1)^j F_\a(\om_1,\om_j)_{\ep_1,\ep_j}
    F_\a(\om_2,\ldots,\widehat{\om_j},\ldots,\om_n)_{\ep_2,\ldots,\widehat{\ep_j},\ldots,\ep_n}
\eeq where the hat over an argument means omission of this
argument. Of course this will be non-zero only for $\sum_{j=1}^n
\ep_j = 0$. Forming the $n\times n$ matrix ${\bf
F}_\a(\om_1,\ldots,\om_n)_{\ep_1,\ldots,\ep_n}$ with matrix
elements $[{\bf
F}_\a(\om_1,\ldots,\om_n)_{\ep_1,\ldots,\ep_n}]_{i,j} =
F_\a(\om_i,\om_j)_{\ep_i,\ep_j}$, the ``multi-particle'' form
factors can be written as Pfaffians\footnote{ Using this, one can
immediately write the form factor expansion \eq{genexp} as a
Fredholm determinant in the case $\a=\a'$. }: \beq
    F_\a(\om_1,\ldots,\om_{n})_{\ep_1,\ldots,\ep_n} =
    {\rm Pf} \lt({\bf F}_\a(\om_1,\ldots,\om_n)_{\ep_1,\ldots,\ep_n}\rt).
\eeq Using \eq{Aacomm}, it is always possible to choose the order
such that $N$ operators with positive $U(1)$ charge are followed
by $N$ operators with negative $U(1)$ charge. Forming the $N\times
N$ matrix ${\bf G}_\a(\om_1,\ldots,\om_N;\,\t\om_1,\ldots
\t\om_N)$ with matrix elements $[{\bf
G}_\a(\om_1,\ldots,\om_N;\,\t\om_1,\ldots \t\om_N)]_{i,j} =
F_\a(\om_i,\t{\om}_j)_{+-}$, we have \beq\lab{mpff}
    F_\a(\om_1,\ldots,\om_{N},\t\om_1,\ldots,\t\om_N)_{\underbrace{+,+,\ldots}_{N},\underbrace{-,-,\ldots}_{N}} =
    (-1)^{N(N-1)/2}\,\det ({\bf G}_\a(\om_1,\ldots,\om_N;\,\t\om_1,\ldots,\t\om_N)).
\eeq Similar expressions are valid for the form factors
$\<0|\Or_\a|k_1,\ldots,k_n\>_{\ep_1,\ldots,\ep_n}$ in the discrete
basis.

\sect{Flat space limit of form factors}

It is a simple matter to verify that the ``two-particle'' form
factors \eq{ffapp} specialize to the known expression in the flat
space limit. Two-particle form factors of the scaling fields
$\Or_\a$ in the flat space limit, with particle at rapidities
$\beta_1$ and $\beta_2$ and states normalized by ${\
}_{\ep_1}\<\beta_1|\beta_2\>^{flat}_{\ep_2} = 2\pi
\delta_{\ep_1,\ep_2} \delta(\beta_1-\beta_2)$, are obtained by
\beqa
&&  \<vac| \Or_\a | \beta_1, \beta_2\>_{+-}^{flat} = \lim_{\mu\to\infty} \times \\
&&  (2\mu)^{\a^2+1} 2\pi \sqrt{\rho(2\mu\sinh(\beta_1))
\rho(2\mu\sinh(\beta_2))
    \cosh(\beta_1)\cosh(\beta_2)} F_\a(2\mu\sinh(\beta_1), 2\mu\sinh(\beta_2))_{+-} \no
\eeqa and the one-point function is obtained by
$$
    \<\Or_\a\>^{flat} = \lim_{\mu\to\infty} (2\mu)^{\a^2}
    \<\Or_\a\>.
$$
Using the expression \eq{ffapp}, this gives the known normalized
flat space form factors, first obtained in \cite{ST}:
$$
    \frc{\<vac| \Or_\a | \beta_1, \beta_2\>_{+-}^{flat}}{
    \< \Or_\a \>^{flat}} =
    \frc{i \sin(\pi\a)}{\cosh\lt(\frc{\beta_1-\beta_2}2\rt)} e^{\a(\beta_1-\beta_2)}.
$$

\end{document}